\documentclass{emulateapj}
\slugcomment{Submitted to the Astronomical Journal, \today}

\shorttitle{SDSS AM CVn Binaries}
\shortauthors{Anderson, Scott F. et al.}

\begin{document}


\title{Ultracompact AM~CVn Binaries from the Sloan Digital Sky Survey:
Three Candidates Plus the First Confirmed Eclipsing 
System\altaffilmark{1}}


\author{Scott F. Anderson\altaffilmark{2},
Daryl Haggard\altaffilmark{2},
Lee Homer\altaffilmark{2},
Nikhil R. Joshi\altaffilmark{2},
Bruce Margon\altaffilmark{3},
Nicole M. Silvestri\altaffilmark{2},
Paula Szkody\altaffilmark{2},
Michael A. Wolfe\altaffilmark{2},
Eric Agol\altaffilmark{2},
Andrew C. Becker\altaffilmark{2},
Arne Henden\altaffilmark{4},
Patrick B. Hall\altaffilmark{5},
Gillian R. Knapp\altaffilmark{6},
Michael W. Richmond\altaffilmark{7},
Donald P. Schneider\altaffilmark{8},
Gregory Stinson\altaffilmark{2},
J.C. Barentine\altaffilmark{9},
Howard J. Brewington\altaffilmark{9},
J. Brinkmann\altaffilmark{9},
Michael Harvanek\altaffilmark{9},
S.J. Kleinman\altaffilmark{9},
Jurek Krzesinski\altaffilmark{9,10},
Dan Long\altaffilmark{9},
Eric H. Neilsen, Jr.\altaffilmark{9},
Atsuko Nitta\altaffilmark{9},
Stephanie A. Snedden\altaffilmark{9}}

%\noindent
\email{anderson@astro.washington.edu}

\altaffiltext{1}{Based on observations obtained with 
%the Sloan Digital Sky Survey and 
the Apache Point Observatory (APO) 3.5m 
telescope, which is owned and operated
by the Astrophysical Research Consortium (ARC).}
\altaffiltext{2}{University of Washington, Department of
   Astronomy, Box 351580, Seattle, WA 98195}
\altaffiltext{3}{Space Telescope Science Institute, 3700 San
	Martin Drive, Baltimore, MD, 21218}
\altaffiltext{4}{US Naval Observatory, Flagstaff Station,
        P.O. Box 1149, Flagstaff, AZ 86002-1149}
\altaffiltext{5}{Department of Physics \& Astronomy, York 
        University, 4700 Keele St., Toronto, ON, M3J 1P3, Canada}
\altaffiltext{6}{Princeton University Observatory, Princeton,
        NJ 08544}
\altaffiltext{7}{Department of Physics,
        Rochester Institute of Technology, Rochester, NY 14623-5603}
\altaffiltext{8}{Pennsylvania State University, Department of 
        Physics \& Astronomy, 525 Davey Lab, University Park, PA 16802}
\altaffiltext{9}{Apache Point Observatory, P.O. Box 59, Sunspot, NM 88349}
\altaffiltext{10}{Mt. Suhora Observatory, Cracow Pedagogical University, 
ul. Podchorazych 2, 30-084 Cracow, Poland}


%\altaffiltext{5}{Patron, Alonso's Bar and Grill}


%\noindent
%\myemail

\begin{abstract}

AM~CVn systems are a rare (about a dozen previously known) class of 
cataclysmic variables, arguably encompassing the shortest orbital periods 
(down to about 10 minutes) of any known binaries. Both binary components 
are thought to be degenerate (or partially so), likely with mass-transfer 
from a helium-rich donor onto a white dwarf, driven by gravitational 
radiation. Although rare, AM~CVn systems are of high interest as possible 
SN~Ia progenitors,
and because they are predicted to be common
sources of gravity waves in upcoming experiments 
such as {\it LISA}. We have identified four new AM~CVn candidates from the 
Sloan Digital Sky Survey (SDSS) spectral database. All four show hallmark 
spectroscopic characteristics of the AM~CVn class: each is devoid of 
hydrogen features, and instead shows a spectrum dominated by helium. 
All four show double-peaked emission, indicative of 
helium-dominated accretion disks. Limited time-series CCD photometric 
follow-on data have been obtained for three of the new candidates from the 
ARC 3.5m; most notably, a 28.3 minute binary period with sharp, deep 
eclipses is discovered in one case, SDSS~J0926+3624. This is 
the first confirmed eclipsing AM~CVn, and our data allow 
initial estimates of binary parameters for this ultracompact system.
The four new SDSS objects 
also provide a substantial expansion of the currently critically-small 
sample of AM~CVn systems. 
\end{abstract}


%% subject header.  Objects should be in the appropriate "individual"
%% headers (e.g. quasars: individual, stars: individual, etc.) with the
%% additional provision that the total number of headers, including each
%% individual object, not exceed six.  The \objectname{} macro, and its
%% alias \object{}, is used to mark each object.  The macro takes the object
%% name as its primary argument.  This name will appear in the paper
%% and serve as the link's anchor in the electronic edition if the name
%% is recognized by the data centers.  The macro also takes an optional
%% argument in parentheses in cases where the data center identification
%% differs from what is to be printed in the paper.

\keywords{binaries: close --- binaries: eclipsing --- novae,cataclysmic 
variables --- white dwarfs --- stars: individual (\objectname{SDSS
J092638.71+362402.4})}


\section{Introduction}

AM~CVn systems are an extremely rare type of cataclysmic
variable with ultrashort binary orbital periods of less than 
about an hour. The most confident current members of this
elite subclass have orbital periods in the range of 
$\sim$10-65 minutes, though two more controversial cases
even display 5-10~minute modulations; AM~CVn systems thus arguably
encompass the shortest orbital periods of any known
class of binaries  \citep[e.g., see reviews 
by][]{war95,nel05}.
%and \citep{nel05}).
These systems are so compact that both binary components are probably
degenerate (or semi-degenerate), likely with mass-transfer from a
helium-rich donor onto a white dwarf, driven by 
gravitational radiation.  Their
optical spectra are distinct from typical cataclysmics: membership in
the AM~CVn class requires a virtual absence of hydrogen lines in their
spectrum, and instead helium lines are prominent---sometimes in
emission, sometimes in absorption. Hydrogen is thought to be lost
from the binary during a prior common-envelope phase, before the
object is seen as an AM~CVn.

The exotic nature of the prototype, AM~CVn (with a 17 minute 
orbit), was realized in the mid-1960's
%(Smak 1967; Paczynzki 1967).  
\citep{sma67,pac67}.
But the passage of another three and
half decades has yielded only about a dozen additional members of this
class; for example, 11 confident plus 2 controversial cases
are discussed in the review by \citet{nel05}, and another 
likely 
addition is reported in a very recent telegram by 
\citet{ryk05} and \citet{wou05}.
%Rykoff 2005, and Woudt \& Warner 2005. 
Although
they have thus far remained elusive, AM~CVn systems have nonetheless 
emerged
as objects of renewed interest for multiple reasons: they may serve as
sites in which to study common envelope binary evolution and unusual
helium-dominated accretion disks; they are possible SN Ia progenitors
%(e.g., Livio \& Riess 2003);
\citep[e.g.,][]{liv03};
and, intriguingly, they are predicted to be one of the most commonly
detected objects in 
upcoming gravity wave experiments like {\it LISA}. 
For example, recent estimates, based on formation and evolutionary models,
suggest that $\sim 10^4$ AM~CVn binaries will plausibly be 
detected/resolved
in gravity waves by {\it LISA} 
%(Nelemans et al. 2004).
\citep{nel04}.

Although most AM~CVn systems were first recognized 
via their optical properties, they may also emit X-rays during
evolutionary phases of substantial mass transfer
%(e.g., Ulla 1995),
\citep[e.g.,][]{ull95},
and are also sometimes detected in the ultraviolet. 
Indeed, two of the shortest-period, but also most controversial,
candidates were noticed first by their X-ray emission 
\citep{mot96,isr99,cro04}.
%(Motch et al. 1996, Israel et al. 1999, Cropper et al.2004).
%\footnote{Note
%that there is discussion whether their 5-10 minute modulations 
%are orbital, and whether or not these two belong to the AM~CVn
%subclass (e.g., Wu et al. 2002, Norton et al. 2004, Cropper et al. 
%2004).}  
The combination of multi-wavelength
data sets can allow for estimates of the mass-transfer rates, 
and provide orbital periods and even their long-term
stability in some cases. Such follow-on studies play an important role 
in establishing the evolutionary stage of individual systems, and in
testing evolutionary models.
High-quality follow-up XMM-Newton spectra
have recently extended study of the AM~CVn unusual-abundance signatures 
out to high-ionization X-ray lines as well 
%(e.g., Strohmayer 2004, Ramsay et al. 2005).
\citep[e.g.,][]{str04,ram05}.

Among several theoretical
formation scenarios for AM~CVn systems 
%(see discussion and references in Nelemans 2005), 
\citep[see discussion and references in][]{nel05},
one popular
model initially involves angular momentum loss in a
double-degenerate white dwarf system due to gravitational-wave
radiation. This leads to a decreasing binary period until, at 
periods of a few minutes mass transfer begins, which 
subsequently widens the binary.
During this interval of mass transfer the system is recognizable as 
an AM~CVn.  The youngest members of the AM~CVn class are (in
this scenario) thought to be those with the highest mass-transfer
rates and shortest periods ($<$20 minutes); they then evolve
to somewhat longer periods (a 20-40 minute period during an
intermediate stage, then a late stage with a 40-60 minute period) as
the binary widens and the mass transfer rate continues
to decrease.
Theoretical models for the
evolutionary relation between orbital period and mass-transfer
rates, in the presence of gravitational radiation, predict a steep 
relation of order $\dot{m} \propto P^{-5}$ \citep{war95}.
%Warner 1995).

Although there have been extensive X-ray and optical 
investigations
of many of the dozen previously-known systems (generally yielding
consistency with such evolutionary models),
confident, global conclusions
remain limited by the current very-small total sample size
of AM~CVn systems. 
The quality data and efficient large-area sky coverage
of the Sloan Digital Sky Survey 
%(SDSS; e.g., York et al. 2000,Abazajian et al. 2005)
\citep[SDSS; e.g.,][]{yor00,aba05}
have proven excellent for finding 
rare objects
including cataclysmic variables 
%(e.g., Szkody et al. 2002, Szkody et al. 2005), 
\citep{szk02,szk05},
and SDSS is now emerging as a potentially major contributor to new 
AM~CVn discoveries in particular.  The first SDSS-selected AM~CVn, SDSS
J1240-0159, was discovered in an early SDSS data release by
%Roelofs et al. (2004, 2005).
\citet{roe04,roe05}.
We describe here a successful search of the SDSS
spectroscopic database for additional AM~CVn
candidates, providing basic data on 4 new candidates---a significant
expansion to this rare subgroup. 

We also describe the
results of our initial time-series ARC 3.5m CCD photometric follow-up
for three of our new candidates. 
Most notably, our 3.5m data reveal the first
example of an eclipsing AM~CVn system,
SDSS~J0926+3624; the latter has a 28.3 minute orbital period
with deep, sharp eclipses.

\section{Four New AM~CVn Candidates from SDSS Spectroscopy}

The SDSS is a multi-institutional
project creating an optical digital imaging and spectroscopic data bank of
a large portion of the celestial sphere, mainly in a region approaching
$\sim10^4$~deg$^2$  centered on the north Galactic polar cap. 
The optical data are obtained by a special purpose 2.5m telescope,
located at Apache Point Observatory, New Mexico, 
equipped with a large-format mosaic camera that can image 
$\sim10^2$~deg$^2$ in 5 colors ($u,g,r,i,z$) in a single night, as well
as a multifiber spectrograph which obtains the spectra of 640 objects
within a 7~deg$^2$ field simultaneously.  The imaging database 
is used to select objects for the SDSS spectroscopic survey, which will
include ($\lambda/\Delta\lambda\sim1800$)
spectrophotometry covering a broad (3800-9200\AA) wavelength regime
for $10^6$ galaxies, $10^5$ quasars, and $10^5$ stars.
Details on SDSS hardware, software, and astrometric, photometric,
and spectral data may be found in a variety of papers, including
\citet{fuk96}, \citet{gun98}, \citet{lup99}, \citet{yor00}, \citet{hog01}, 
\citet{sto02}, \citet{smi02}, \citet{pie03}, and \citet{ive04}. 
%Fukugita et al. (1996), Gunn et al. (1998), Lupton et al. 
%(1999), York et al. (2000), Hogg et al. (2001), Stoughton et al. 
%(2002), Smith et al. (2002), Pier et al. (2003), and Ivezi\'c et al. 
%(2004).
A description of the latest SDSS Public Data Release is given
by \citet{aba05}.
%Abazajian et al. (2005).

As part of the efforts of one of the SDSS scientific working groups (the 
``Serendipity
Working Group"), we are engaged in a visual examination of 
a large number of SDSS spectra. A recent perusal includes 
examination of 286,000 SDSS object spectra from 470 distinct 
SDSS spectroscopic  plates, encompassing of order 2400 deg$^2$ of sky. 
This visual search has 
found/recovered several thousand spectroscopically unusual objects, 
with many in the SDSS collaboration contributing their expertise to
classify and understand the unusual spectra, and with identifications ranging
from unusual white dwarfs to extreme broad absorption line
quasars. Each of the 4 new SDSS AM~CVn candidates reported
herein was found in the course of this specific recent visual search
for spectroscopically unusual objects, and each AM~CVn candidate
shows a hallmark helium-dominated optical spectrum; all four 
new SDSS candidates,
SDSS~J012940.05+384210.4,
SDSS J092638.71+362402.4,
SDSS J141118.31+481257.6, and SDSS J155252.48+320150.9,
have helium in emission (Figure 1). Hereafter, we will 
usually refer to these via the shorter nomenclature, SDSS~J0129+3842,
SDSS J0926+3624, SDSS J1411+4812, and SDSS J1552+3201.
Of course, it is their distinctive spectra which attracted
attention to these candidates in this visual search of SDSS
spectra.
\begin{figure*}[!ht]
\epsscale{1.02}
\plotone{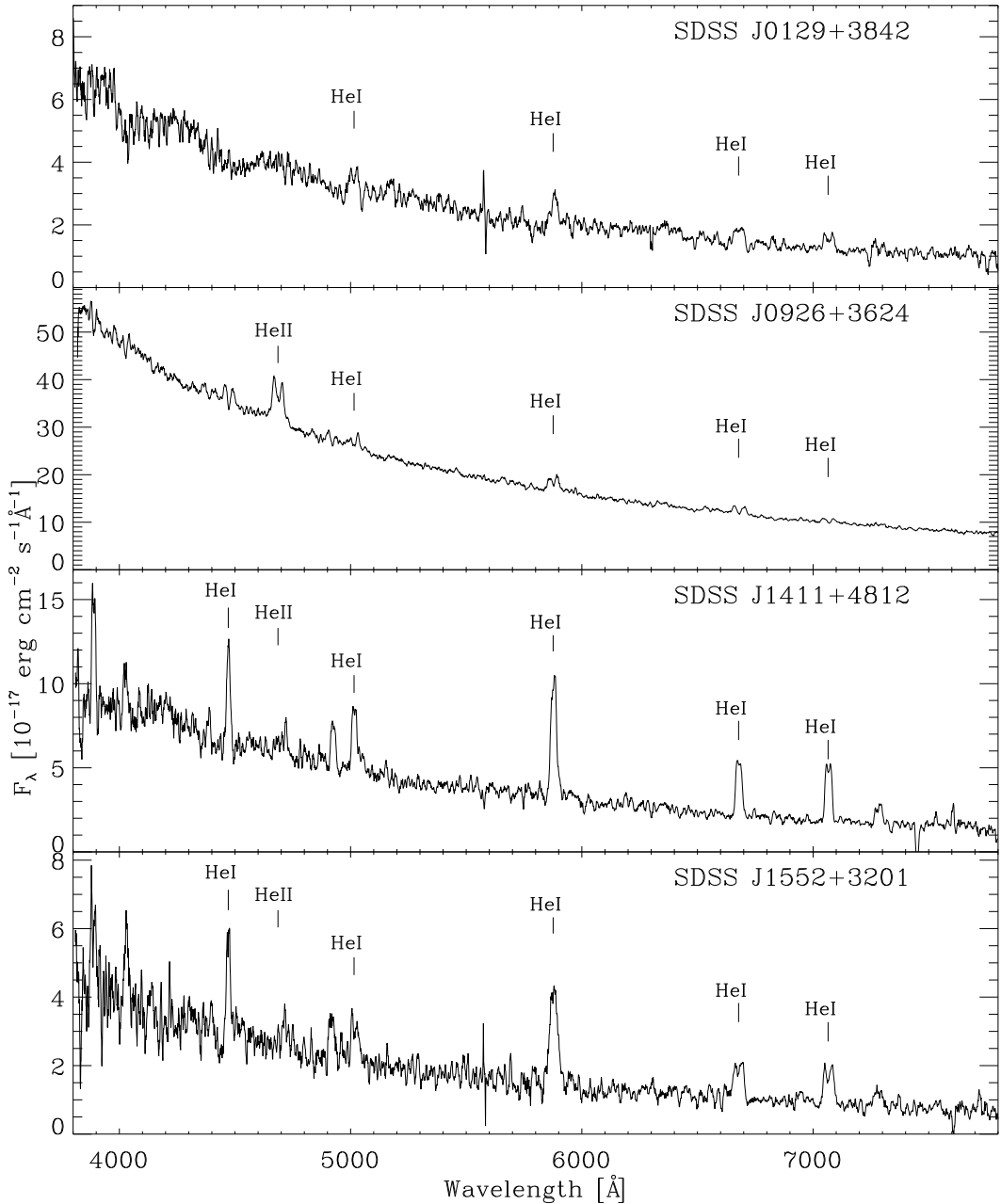}
\vspace*{1cm}
\caption{SDSS discovery spectra of the four new AM~CVn candidates,
displayed here with a 5-point boxcar smooth.
Each candidate was discovered first in the course of a visual examination 
of several hundred thousand SDSS spectra.
Each spectrum is dominated by helium, and each
shows double-peaked emission indicative of a helium accretion disk.
\label{fig1}}
\end{figure*}

On the other hand, such visual searches are potentially highly biased and 
incomplete, and so we have also examined a preliminary
version of the SDSS Data 
Release~4 (DR4) spectroscopic database via a more quantitative
algorithmic search; this
includes just over 1050 spectroscopic plates 
covering about 4700 square degrees, and including 640,000 object spectra. 
%\footnote{
(Though note that some ``special plates" added very recently to DR4 are 
not included.) 
We queried the DR4 
database to return a list of spectra in 
which the SDSS pipeline data reduction algorithms 
%(Stoughton et al. 2002) 
\citep{sto02} found any emission line centered within 20~\AA\  
wavelength-windows around  HeII $\lambda$4686 or HeI $\lambda$5875;
these were the highest equivalent-width helium lines 
in the four AM~CVn candidates found in the visual search. 
The queries 
returned cases in which the pipeline Gaussian-fits to HeI or HeII 
equivalent 
widths exceeded 3\AA, if either of the two lines was detected.
The algorithmic search is also 
limited to emission-line objects,
but at least represents a quantitative and uniform search.
\begin{deluxetable*}{lcccccccccc}
%\rotate
\tabletypesize{\footnotesize}
\tablecaption{Basic SDSS Data on the Four New AM~CVn Candidates}
\tablewidth{0pt}
\tablecolumns{11}
                                                                                                                                                       
\tablehead{ \colhead{RA,Dec name} &
   \colhead{{\it u}} & \colhead{{\it g}} &
   \colhead{{\it r}} &
   \colhead{{\it i}} & \colhead{{\it z}} &
   \colhead{line} &
   \colhead{EW} &
   \colhead{FWHM} \\
   \colhead{SDSS} & 
   \colhead{ } & \colhead{ } &
   \colhead{ } & \colhead{ } & \colhead{ } &
   \colhead{ } &
   \colhead{(\AA)} &
   \colhead{(km s$^{-1}$)}}
                                                                                                                                                       
\startdata
J012940.06+384210.4 & 19.66 & 19.81 & 20.04 & 20.23 & 20.60 & HeI 5875 & 23 & 1300 \\
J092638.71+362402.4 & 18.71 & 18.97 & 19.19 & 19.39 & 19.38 & HeII 4686 & 13 & 3400 \\
J141118.24+481257.6 & 19.30 & 19.37 & 19.51 & 19.73 & 19.80 & HeI 5875 & 65 & 1400 \\
J155252.48+320150.9 & 20.16 & 20.24 & 20.32 & 20.41 & 20.54 &  HeI 5875 & 75 & 2200 \\
\enddata
                                                                      
\end{deluxetable*}

The algorithmic SDSS spectral database queries initially returned about
19000 SDSS spectra potentially having emission
at wavelengths of interest for AM~CVn binaries. Each of these
algorithmically-chosen spectra was then further examined
by eye to ascertain a rough spectral type. Many
were quasars or emission-line galaxies, redshifted such that 
their emission lines fall within the wavelengths
of the search windows around the HeI or HeII lines. 
Three SDSS AM~CVn candidates initially discovered by
other means fall within the DR4 spectroscopic survey
area: SDSS~J0129+3842 and SDSS~J0926+3624 found initially
in the visual perusal discussed above and reported
here for the first time, as well as the previously published 
case, SDSS~J1240-0159. All three were successfully
recovered via our algorithmic search, but
no additional (spectroscopically)
compelling AM~CVn candidates emerged from the
algorithmic search of DR4 SDSS spectra.

The predominance of helium emission in each 
spectrum displayed in Figure 1 alone argues 
strongly for an AM~CVn classification. 
Prominent in most objects are He I 3888\AA, 4026\AA, 4471\AA, 
5015\AA, 5875\AA, 6678\AA, 7065\AA, and 7281\AA; but for SDSS~J0926+3624,
HeII $\lambda$4686 is also very strong.
In addition, all four new AM~CVn candidates 
show double-peaked helium emission, very likely arising from
(helium-dominated) accretion disks.
%\footnote{
In some cases, there is evidence for a third 
additional, central emission ``spike", as has been
previously noted in some other AM~CVn systems like SDSS~J1240-0159 
%(e.g., Roelofset al. 2005), 
\citep[e.g.,][]{roe05},
but additional spectroscopy is needed to verify.
Basic optical SDSS
astrometric, photometric, and selected emission line data 
(for the strongest helium emission line) 
are provided in Table~1. 
Of course, these AM~CVn candidates
are anticipated (and confirmed in some cases)
to be variable, so
the SDSS photometry is only 
representative of the epoch of SDSS images.

The visual search of SDSS spectra that originally
revealed these four new AM~CVn candidates was done without restriction
on how the object was initially chosen as an SDSS spectroscopic target.
But in summary it is mainly the blue-object color selection
SDSS ``serendipity" algorithms, and secondarily
the ``white-dwarf"  or ``hot-standard star"  
SDSS algorithms that have yielded the new AM~CVn candidates.
Both the SDSS~J1411+4812 and 1552+3201 spectra, as well as that of
the original \citet{roe04}
%Roelofs et al. 2004 
object SDSS~J1240-0159, were obtained
as SDSS serendipity targets. SDSS~J0129+3842 was targeted under an early
version of white-dwarf target selection for the SEGUE 
%(e.g., Beers et al. 2004) 
\citep{bee04}
extension of SDSS. 
SDSS~J0926+3624 was actually targeted
for a spectrum as an object potentially suitable
as a hot, spectrophotometric standard star, though it was also selected as 
a possible
spectral target by both serendipity and white-dwarf algorithms.
These AM~CVn candidates (not too surprisingly)  occupy blue regions of
SDSS color-color diagrams that also include
white dwarfs (Figure 2).

\begin{figure}
\voffset=-0.5in
\epsscale{1.0}
\centerline{\plotone{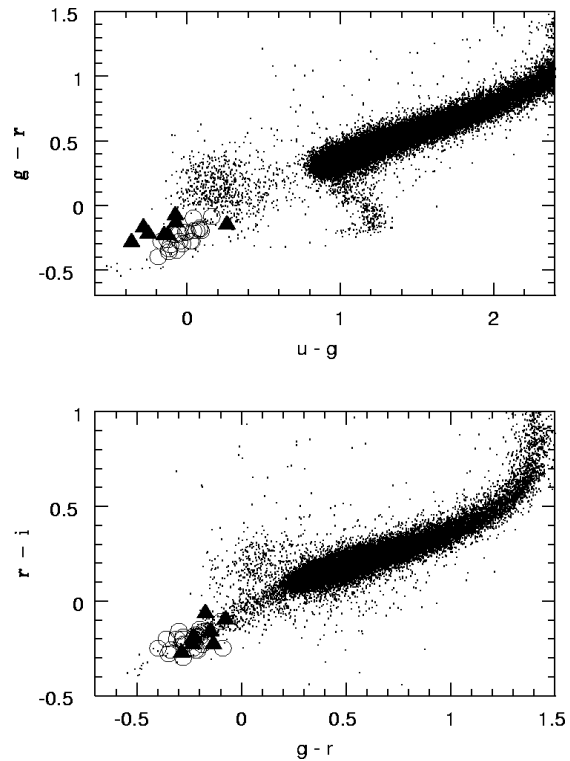}}
\voffset=-0.5in
\caption{Representative SDSS color-color diagrams for AM~CVn
systems and other SDSS stellar objects. The small black points show the 
colors of stellar objects having 
reliable photometry, derived from a random 100~deg$^2$ of SDSS. 
Over-plotted are SDSS colors
of AM~CVn candidates/confirmations (solid, black triangles);
the triangles show the colors of the 4 new AM~CVn SDSS candidates
described here, plus those of
four other previously-known AM~CVn systems that also fall within
the SDSS DR4 {\it imaging} area.
(Note: the prototype, AM~CVn, is not shown
due to its bright magnitude which renders its SDSS photometry
uncertain).
For additional color-comparisons, SDSS DB white dwarfs from 
%Harris et al. (2003) 
\citet{har03} are also plotted (open circles).
The SDSS colors of these AM~CVn systems are roughly
similar to those of white dwarfs.
\label{fig2}}
%\end{minipage}
%\end{center}
\end{figure}

\section{Followup Lightcurves, and Confirmation of the First Eclipsing 
AM~CVn}

In follow-up observations, we used the SPIcam CCD imager
on the ARC 3.5m on two partial nights
in spring 2005 to observe the three of our 
new AM~CVn candidates with accessible RAs.
We used non-standard on-chip binning, along with customized
subframes, to reduce deadtime during individual read outs.
A short time-series sequence (spanning about 1 hour, limited
by weather) was obtained for 
SDSS~J0926+3624 on UT 2005 March 16. 
On UT 2005 April 13, longer time-series were 
obtained for both SDSS~J0926+3624 and SDSS~J1411+4812, 
with a short time-series sequence for SDSS~J1552+3210.
Neither night was photometric, but differential photometry
was obtained, relative to several slightly brighter comparison stars 
included in each
image. We used the SPIcam SDSS $g$-band filter in all
cases (as uncooperative
weather did not allow sufficient time to obtain time-series
in additional filters).

For SDSS~J1411+4812, we collected a CCD time-series spanning
3~hours on our April night, and comprised of 
mainly 40~sec individual exposures,
yielding somewhat better than 60~sec time-resolution (e.g., with a 
deadtime of about 18~sec between each 40~sec exposure).
Application of Lomb-Scargle analysis \citep{lom76,sca82} 
%(Lomb 1976; Scargle 1982) 
to these time-series 
data reveals no significantly-detected periodic modulation for 
SDSS~J1411+4812. Using Monte Carlo simulations
to quantify, we place an upper limit on any sinusoidal modulation of the
lightcurve for SDSS~J1411+4812 of $<2 \%$ (semi-amplitude) on
timescales of 4 to 60 minutes, at the epoch of our 3.5m observations.
However, the 3.5m lightcurve for SDSS~J1411+4812 does suggest non-periodic
variations of typical amplitude $\sim 0.05$ mag.
\begin{figure*}[!ht]
\leavevmode
\begin{center}
\includegraphics[angle=0,scale=0.72]{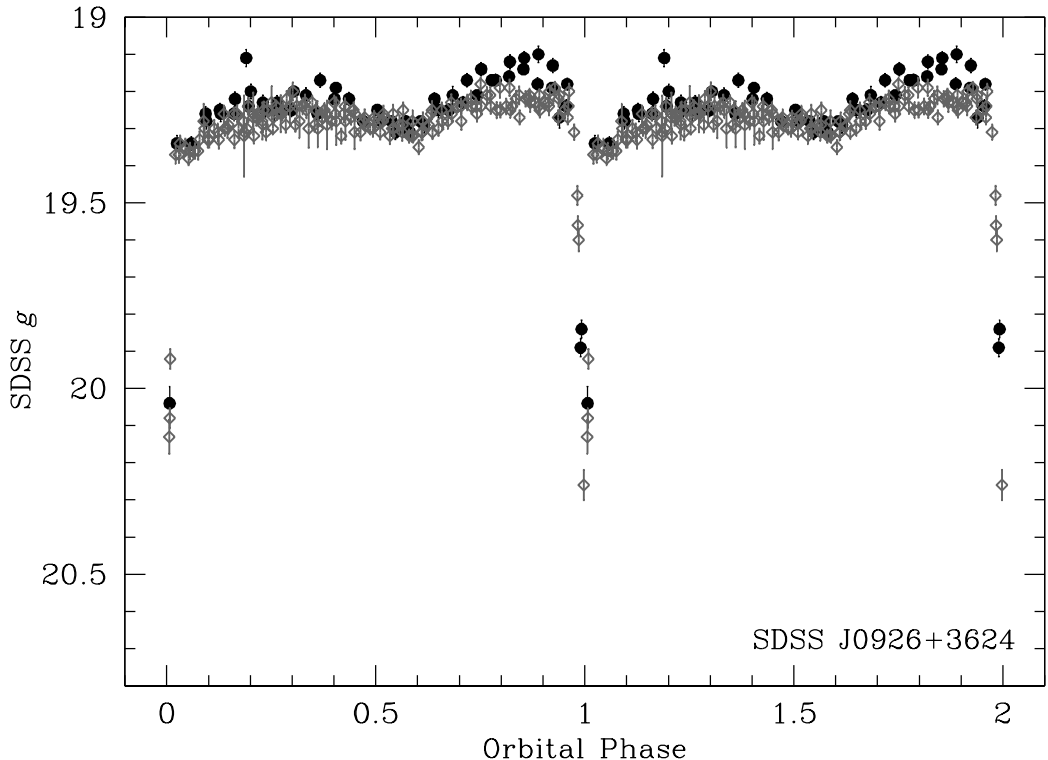}
\caption{Folded $g$-band (differential) lightcurve data for SDSS 
J0926+3624, obtained using
SPIcam on the ARC 3.5m, during a portion of two nights in 
March (solid, black circles) and April (open, grey diamonds) of 2005. 
Sharp, 
deep eclipses with
a period of 28.3 minutes are evident even in the raw lightcurve, and confirmed
with Lomb-Scargle analysis; the data are folded here on the best-fit 
28.31 
minute
period. These follow-on 3.5m data confirm that SDSS~J0926+3624 is an 
ultracompact binary, and the first example of an eclipsing AM~CVn.
\label{fig3}}
\end{center}
\end{figure*}

For SDSS~J1552+3204, we squeezed in a 1.5 hour time-span sequence on our April
night with the ARC 3.5m, taking  60~sec individual exposures (with about 
80~sec time-resolution). 
We also had obtained an earlier time-series CCD sequence
for SDSS~J1552+4203, using the Naval Observatory (NOFS)
1.0m on UT 2004 June 12; a 4-hour time-span was
covered with about 5-minute time-resolution in unfiltered light.
Lomb-Scargle analysis reveals a possible very weak, 0.04~mag 
semi-amplitude modulation with a period of about 56 minutes 
in our 3.5m data, or perhaps 
half that period in the NOFS data. Monte Carlo 
analyses suggest a marginal significance of $\sim99\%$ for this possible 
periodicity,  though the low
amplitude and significance, and the brief cycle-count of the 3.5m 
observations, 
demand additional data to verify.

For SDSS~J0926+3624, both the 1-hour timespan March sequence with 40~sec 
exposures (and 60~sec time-resolution),
and the longer 2-hour timespan April sequence with 20~sec exposures
(and 40~sec time-resolution) reveal a strongly 
detected 28.31$\pm$0.01 minute modulation, with striking eclipses 
(Figure~3).
From our April data, we estimate eclipse centers at
${\rm HJD(TT)} = 2453473.725393(3) + E\times0.01966(1)$. The eclipses are sharp 
and deep, with mid-eclipse depth of (at least) about 1 magnitude.
A Gaussian fit to the sharp portion of the eclipse duration yields a 
full-width at half minimum of about 40~seconds 
and, for example, a $\sim1.3$ minute duration full width at
5\% of the minimum eclipse depth. As our time-resolution is only of order
40~sec, the eclipses could be even sharper/deeper.
Application of Lomb-Scargle analysis confirms this (obvious) period
with very high confidence (e.g., $>99.9\%$ significant in the April data). 
The lightcurve also shows additional
reproducible structure that is seen in both March and April 2005
3.5m data; for example, there is
a plateau just following the eclipse,
and note the low-amplitude relative depression
near phase 0.6.
As far as we are aware, SDSS~J0926+3624 is the 
first example of an eclipsing AM~CVn.

\section{Discussion}

Our visual spectroscopic search yielded 4 new AM~CVn candidates
from recent SDSS spectroscopic plates
covering about 2400 deg$^2$, while the subsequent algorithmic search 
recovered 3 AM~CVn candidates (two of the new candidates 
discussed herein, as well as the Roelofs et
al. 2004 object, SDSS~J1240-0159) from plates encompassing 
about 4700 deg$^2$.
We are not aware of any examples in the SDSS spectral database of 
strong emission-line AM~CVn binaries
obviously missed by either visual or algorithmic searches.
%That is, the algorithmic approach applied to the DR4 spectral database 
%recovers all the cases found in the visual search (and SDSS 
%J1240-0159), but no additional cases. 
On the other hand, the overlap-numbers are too small to be considered 
definitive.
Accounting for  the 250 plates covered both in the visual and 
algorithmic searches, the combined current SDSS spectroscopic
yield is 5 (emission-line) AM~CVn candidates from a region 
encompassing about 5900 deg$^2$ of sky. We thus estimate that 
spectroscopically similar
AM~CVn systems might be expected to be found from the SDSS 
spectral database at a rough
surface density of order 1 every 1200 deg$^2$. 

Estimates of the actual surface density of SDSS AM~CVn binaries (even just 
those with helium in emission) are complicated in that
the AM~CVn candidates found thus
far with SDSS were chosen for spectroscopy by: (i) several different target
selection algorithms, having differing color-selection criteria and 
differing limiting magnitudes;  and, (ii) selection algorithms (mainly) 
that in practice receive a spectroscopic fiber only when the main SDSS 
galaxy and quasar surveys do not consume all available fibers for a given
spectroscopic plate. Because of such complications, we more conservatively
note that
enough spare spectroscopic fibers (e.g., for ``serendipity" targets)
generally have not been available to fully sample all potentially 
interesting
candidates having SDSS colors similar to those of the candidate/confirmed
AM~CVn cases. For example, if considering a small box in
$u-g$, $g-r$, $r-i$ multicolor-space that barely encompasses the
bulk of candidate/confirmed AM~CVn binaries in DR4 imaging (see Figure 2), 
only of order one-fifth of the similar-color objects cataloged in SDSS DR4 
images with $15<g<20.5$ also have available SDSS DR4 
spectroscopy. Spectroscopic incompleteness of $>80\%$ (or even much 
larger) is thus not implausible, and even just for AM~CVn candidates 
having spectral characters similar to those presented here.

Our spectroscopic recognition (both visual and algorithmic) 
of AM~CVn candidates reported
here relies on the detection of helium emission lines, and thus it
is not surprising that the current SDSS sample
includes no confirmed examples of the shortest-period ``high-state" systems
like AM~CVn itself, which show helium predominantly in absorption. 
On the other hand, the predominance of strong helium emission, and the 
low-amplitude 
(less than about 0.1 mag) variability found for 
SDSS~J1411+4812 and SDSS~J1552+3201 on both long- 
and short-terms, are consistent
with their being ``low-state" AM~CVn systems, perhaps with 
orbital periods in
the range of about 40-60 minutes. The marginal 56~minute
low-amplitude photometric periodicity for SDSS~J1552+3201, of course,
would also be consistent with this notion; but we emphasize 
again that this possible
periodicity should not be considered secure, pending 
further observations. Time-resolved optical spectroscopy may well be 
the preferred test of ultracompact binarity in such cases, as it has also
been for several other possibly-similar ``low-state" AM~CVn systems
including SDSS~J1240-0159 (e.g., see Roelofs et al.
2005, or broad discussion in Nelemans 2005).

SDSS~J0926+3624 on the other hand displays long-term 
variability (e.g., having varied by about a magnitude
between the epochs of SDSS spectra and SDSS
images), as well as the striking 28.3 minute
eclipsing lightcurve reported herein. Both the
short orbital period, and the longer-term
variability, suggest that SDSS~J0926+3624 might be an AM~CVn at
an intermediate evolutionary stage.
% (e.g., like CP Eri).
Of course quality, time-resolved
spectra (in both high and low-states) are another obvious
imperative given the eclipsing nature demonstrated in this paper.
As the first example of an eclipsing AM~CVn,
SDSS~J0926+3624 provides an excellent potential 
opportunity to empirically
confirm binary parameters that may be less directly
obtained in the other AM~CVn systems. 

It is interesting to note that AM~CVn binary parameters, 
theoretically preferred in some models and/or empirically
inferred for some other AM~CVn
systems, are at least not inconsistent with 
our initial SDSS and 3.5m observations and their 
implications for SDSS~J0926+3624.
% considered here in approximate fashion. 
For example, the observed velocity (full) width of the HeII $\lambda$4686 
line near the continuum level
is $\sim$6100~km s$^{-1}$; under the assumption
that $\sin~i$ is approximately unity in this eclipsing case,
this implies
a compactness parameter for the accreting primary of
$M_1/R_1 \sim 50 M_\odot/R_\odot$, if a Keplerian disk extends
to near the primary surface (or half that value for freely 
falling material). For a zero-temperature white dwarf mass-radius
relation \citep[e.g.,][]{rap84,nau72}, 
%(e.g., Rappaport and Joss 1984, Nauenberg 1972), 
this 
Keplerian-disk assumption in turn then
implies that the accreting primary
has $M_1\sim0.6M_\odot$ and 
%$R_1\sim0.012 R_\odot$.
$R_1\sim 9000$~km.
If the HeII disk emission region does not extend down to the
vicinity of the primary white dwarf surface, then $M_1$ would be
larger (and $R_1$ smaller) than these representative values.

Assuming that the secondary fills its Roche lobe (and adopting
an approximate expression for the volume-equivalent 
Roche lobe radius from \citet{pac71} 
%Paczynski 1971, 
appropriate in the case 
of $M_2/M_1<0.8$), Kepler's third law also implies that the secondary
mass and radius are approximately related via
$R_2\approx 0.234 (P/60~{\rm min})^{2/3}M_2^{1/3}$, where
$R_2$ and $M_2$ are both in solar units in the latter expression.
If the secondary in SDSS~J0926+3624 is a degenerate helium
white dwarf as is often suggested for AM~CVn systems, then
again invoking a zero-temperature, fully-degenerate mass-radius relation,
also provides an estimate of the secondary parameters: 
$M_2\sim0.02M_\odot$ and $R_2 \sim 25000$~km. For such a 
%0.018,0.036
system with binary separation $\sim 1.8\times10^5$~km, the relative 
transverse velocities of the secondary
and primary (of course, predominantly due to the motion of the former) at
mid-eclipse would then be $v_{rel} \sim 670$~km s$^{-1}$, and the sharp 
portion of the eclipse duration might be expected 
to be of order $2R_2/v_{rel}\sim 75$~sec
(e.g., if the sharp-portion is mainly an eclipse by the secondary
of a relatively small region, such as a hot spot). The approximate 
agreement of such a simple model with the observed duration of 
about 65-85s for the sharp portion of the eclipse in SDSS~J0926+3624
is interesting. Further modeling of high time resolution eclipse
data may allow constraints to be placed on non-zero temperature
mass-radius relations, with various levels of degeneracy 
%(Deloye \& Bildsten, 2003; Deloye, Bildsten, \& Nelemans 2005).
\citep{del03,del05}.

The sharpness of the eclipses in SDSS~J0926+3624 also suggests 
the long-term possibility of measuring the change in the binary orbital
period, thereby constraining the mass transfer rate and 
aspects of gravitational radiation. For example,
\citet{str05} and \citet{isr04} have successfully 
obtained
such information via measures of the period derivative of
the 5~minute modulations in RX~J0806.3+1527 (which has also 
sometimes been argued to be an AM~CVn system). Accounting 
for  the loss of angular momentum by gravitational waves in a system 
with two degenerates, plus conservative mass transfer via Roche-lobe 
overflow,
the anticipated period change 
%(e.g., Taylor \& Weisberg 1982) 
\citep[e.g.,][]{tay82}
for SDSS~J0926+3624 should be 
of order $\dot P \sim 3\times 10^{-13}$.
This might be expected to lead to a deviation from a
linear $O-C$ eclipse-timing curve of about 1 second over a timescale of 
three years, a value which may be detectable with precise timing
%(e.g., Paczynski 1967).
\citep[e.g.,][]{pac67}.
The implied accretion rate is about $4\times 10^{-11} M_\odot$ yr$^{-1}$
(assuming adiabatic, conservative mass transfer and a zero-temperature
equation of state for the secondary star).
At a fiducial distance of $<$0.5~kpc, the expected gravitational wave 
strain is $h > 3\times
10^{-23}$, which is similar to the $1\sigma$ detection threshold of 
{\it LISA} 
%(e.g., see recent discussions about AM~CVn binaries and {\it LISA} in 
%Nelemans et al. 2004 or Strohmayer 2005).
\citep[e.g., see recent discussions about AM~CVn binaries and {\it
LISA} in][]{nel04,str05}.

In summary, along with the original SDSS AM~CVn found by 
%Roelofs et al. (2004, 2005), 
\citet{roe04,roe05}, the 4 additional new SDSS finds presented here
provide a yield of 5 SDSS AM~CVn candidates thus far; this is a substantial
addition to the elite AM~CVn subclass, compared to the dozen other
cases previously known. Two of these five SDSS objects
are also now strongly confirmed as ultrashort-period binaries 
(SDSS~J1240-0159 by Roelofs et al. 2005,
and J0926+3624 in this paper). SDSS~J0926+3624 reported here is the first 
confident example of an eclipsing
AM~CVn. Our initial approximate considerations presented above
for SDSS~J0926+3624
presage that future detailed modeling of the 28.3 minute 
eclipsing lightcurve and its long-term temporal stability, 
double-peaked spectral line profiles, and follow-on radial 
velocity studies and multiwavelength observations, should 
provide an excellent empirical testbed of various models for AM~CVn 
systems.

\acknowledgments

    Funding for the creation and distribution of the SDSS Archive has been provided by the 
Alfred P. Sloan Foundation, the Participating Institutions, the National Aeronautics and 
Space Administration, the National Science Foundation, the U.S. Department of Energy, the 
Japanese Monbukagakusho, and the Max Planck Society. The SDSS Web site is 
http://www.sdss.org/.

    The SDSS is managed by the Astrophysical Research Consortium (ARC) for the 
Participating Institutions. The Participating Institutions are The University of Chicago, 
Fermilab, the Institute for Advanced Study, the Japan Participation Group, The Johns 
Hopkins University, the Korean Scientist Group, Los Alamos National Laboratory, the 
Max-Planck-Institute for Astronomy (MPIA), the Max-Planck-Institute for Astrophysics 
(MPA), New Mexico State University, University of Pittsburgh, University of Portsmouth, 
Princeton University, the United States Naval Observatory, and the University of 
Washington.

   P. Szkody acknowledges support from NSF AST-0205875. We thank Hugh 
Harris for useful discussions.

\end{document}